\documentclass[preprint]{elsarticle}

\usepackage{lineno,hyperref}
\usepackage{balance}
\usepackage{amsmath}
\usepackage{amssymb}
\usepackage{mathrsfs}
\usepackage{graphicx}
\usepackage{epstopdf}
\usepackage{bm}
\usepackage{color}
\usepackage{subfig}

\modulolinenumbers[5]

\journal{Scientific Report}

\newcommand{\Rt}{\mathcal{R}_t}
\newcommand{\Rn}{\mathcal{R}_0}

\begin{document}
	
	\begin{frontmatter}
		
		\title{A new estimation method for COVID-19 time-varying reproduction number using active cases}
		\tnotetext[mytitlenote]{Corresponding author. Email Address: agus.hasan@ntnu.no (A. Hasan).}
		
		\author[A]{A.\ Hasan*}
		\author[B,B1]{H.\ Susanto}
		\author[B2]{V.R.\ Tjahjono}
		\author[C]{R.\ Kusdiantara}
		\author[B2]{E.R.M.\ Putri}
		\author[D]{\;\;P.\ Hadisoemarto}
		\author[C]{N.\ Nuraini}
		
%		\address[A]{Center for Unmanned Aircraft Systems, University of Southern Denmark, Denmark}
		\address[A]{Department of ICT and Natural Sciences, Norwegian University of Science and Technology, Alesund, Norway}
		\address[B]{Department of Mathematics, Khalifa University, PO Box 127788, Abu Dhabi, UAE}
%		\address[B1]{Department of Mathematical Sciences, University of Essex, United Kingdom}
		\address[B2]{Department of Mathematics, Institut Teknologi Sepuluh Nopember, Surabaya, Indonesia}
		\address[C]{Department of Mathematics, Institut Teknologi Bandung, Bandung, Indonesia}
		\address[D]{School of Medicine, Universitas Padjadjaran, Sumedang, Indonesia}
		
		\begin{abstract}
			We propose a new method to estimate the time-varying effective (or instantaneous) reproduction number of the novel coronavirus disease (COVID-19). The method is based on a discrete-time stochastic augmented compartmental model that describes the virus transmission. A two-stage estimation method, which combines the Extended Kalman Filter (EKF) to estimate the reported state variables (active and removed cases) and a low pass filter based on a rational transfer function to remove short term fluctuations of the reported cases, is used with case uncertainties that are assumed to follow a Gaussian distribution. Our method does not require information regarding serial intervals, which makes the estimation procedure simpler without reducing the quality of the estimate. We show that the proposed method is comparable to common approaches, e.g., age-structured and new cases based sequential Bayesian models. We also apply it to COVID-19 cases in the Scandinavian countries: Denmark, Sweden, and Norway, where the positive rates were below 5\% recommended by WHO.
		\end{abstract}
		
%		\begin{keyword}
%			COVID-19, estimation theory, reproduction number.
%		\end{keyword}
		
	\end{frontmatter}
	
	%\linenumbers
	
	\section*{Introduction}
	
	The coronavirus disease 2019 (COVID-19), a disease outbreak of atypical pneumonia that originated from Wuhan, China \cite{ncov20}, has caused globally at least 250 million confirmed cases, including an estimated 5 million deaths in approximately 221 countries and territories by November 2021. The World Health Organization (WHO) declared the COVID-19 crisis a pandemic on 11 March 2020.
	
	In modelling the disease's transmission as well as to inform and evaluate control policies, it is particularly important to estimate its reproduction number. Early estimates for COVID-19 basic reproduction number $\Rn$, that denotes the transmission potential of infectious disease when introduced to a completely susceptible population, ranged from 1.4 to 6.49 \cite{yliu20}. The effective (or instantaneous) reproduction number $\Rt$, on the other hand, reflects the extent of transmission in the presence of population immunity or intervention. Thus, the estimation of $\Rt$ is important for evaluating public measure success. However, estimation of $\Rt$ is sensitive to the model structure and parameter assumptions \cite{dela19}. As a case in point, due to incorporation of more individual case information and travel data, the estimate for $\Rn$ in Wuhan was revised upward from 2.2-2.7 to 5.7 \cite{sanc20}. On the other hand, data inavailability or poor quality often hinders the use of certain estimation methods, such as serial interval data that are usually needed to estimate $\Rt$ (e.g., Fraser \cite{fras07}, Wallinga and Teunis \cite{wall04}, Cauchemez et al.\ \cite{cauc06}, White and Pagano \cite{whit08}).
	
	In the course of calculating the exact value of $\Rt$, especially when the data has not yet reached its peak, precise assumptions and data estimates are needed. Nishiura et al.\ \cite{nish10} discussed a likelihood-based approach to estimate $\Rt$ from early epidemic growth data, while Cazelles at al. \cite{bernard18} used stochastic models for the disease dynamics coupled with particle Markov chain Monte Carlo algorithm. Using the compartmental Susceptible-Infectious-Recovered (SIR) model, Bettencourt and Ribeiro \cite{bett08} use the incidence data to estimate $\Rn$ and $\Rt$. In this paper, based on the Susceptible-Infectious-Recovered-Dead (SIRD) model as a reference, we develop a novel approach to estimate $\Rt$ of COVID-19. It uses information on the number of infected or active ($I$), recovered ($R$), and death ($D$) cases, which are readily available for all affected countries, so that they can be accessed rather easily. This method does not require information regarding serial intervals, which makes the estimation procedure simpler without reducing the quality of the estimate. We assume mass population testing is sufficiently enough, such that the positive rate is below 5\% recommended by WHO. This is to ensure data quality of the number of infection is acceptable since asymptomatic carrier transmission is often underestimated \cite{Aaron}.
	
	The reproduction number is estimated from reported cases under uncertainties using a two-stage estimation method based on the Extended Kalman Filter (EKF) and a low-pass filter. The method not only considers the nominal number of reported cases, but also its daily pattern. To show our method's practical ability, we apply it to COVID-19 cases in the Scandinavian countries: Denmark, Sweden, and Norway, and compare the results with two commonly used Bayesian methods due to Bettencourt and Ribeiro \cite{bett08} and Cori et al. \cite{fras07,cori13}. We show that the results are indeed comparable. Remark that a similar approach, developed independently, can be found in \cite{Francisco}. The difference is in computational technique to estimate the reproduction number. In this paper, we estimate the reproduction number using EKF, while in \cite{Francisco} it was estimated using Kalman smoother.
	
	%Our paper is structured as follows. In Section 2, we discuss the mathematical model that will base the method. We then discuss the two-stage estimation method in Section 3. In Section 4 we apply the method to estimate the effective reproduction number of COVID-19 in Denmark, Sweden and Norway. We also compare the results with $\Rn$ and $\Rt$ calculated using the methods of \cite{bett08} and \cite{cori13}, respectively. We conclude our work in Section 5.
	
	\section*{A Discrete-Time Stochastic Augmented Compartmental Model}
	
	Our estimation method is based on the compartmental SIRD model that can be written as the following first-order nonlinear differential equations:
	\begin{eqnarray}
		\dot{S}(t) &=& -\frac{\beta I(t)S(t)}{N},\label{Sc}\\
		\dot{I}(t) &=& \frac{\beta I(t)S(t)}{N}-(\gamma+\kappa) I(t),\label{Ic}\\
		\dot{R}(t) &=& \gamma I(t),\\
		\dot{D}(t) &=& \kappa I(t),\label{Dc}
	\end{eqnarray}
	where $S$, $I$, $R$, and $D$ 
	denote the number of susceptible cases, the number of active cases, the number of recovered cases, and the number of deceased cases, respectively. $N$ is the total number of population, $\beta$ is the average number of contacts per person per time, while $\gamma$ and $\kappa$ are the recovery and death rate. Remark that the value of $\beta$ is time-varying due to intervention, i.e., $\beta$=$\beta(t)$. To use the model, we require information on the average infectious time $T_i$ and the Case Fatality Rate (CFR), so that
	\begin{eqnarray}
		\gamma = \frac{1-\text{CFR}}{T_i},\,
		\kappa = \frac{\text{CFR}}{T_i}.
	\end{eqnarray}
	For COVID-19, we take $T_i=9$ as the infectious period on average lasts for 9 days (7-11 days with 95\% CI) \cite{Lan20}, while the CFR is assumed around 1\%. The time-varying effective reproduction number is then given by:
	\begin{eqnarray}
		\Rt(t) = \frac{S(t)}{N} \left(\frac{\beta(t)}{\gamma+\kappa}\right)\approx\frac{\beta(t)}{\gamma+\kappa}.\label{RtD}
	\end{eqnarray}
	The approximation is under the assumption that government intervention is taken at an early stage so that the susceptible is relatively the same over time as the total population. This is the case especially for emerging diseases. 
	We modify the SIRD model by augmenting the following two equations into the system:
	\begin{eqnarray}
		\dot{E}(t) &=& (\gamma+\kappa)I(t)-E(t),\quad
		\dot{\Rt}(t)=0. \label{au}
	\end{eqnarray}
	The former equation takes into account the daily number of new reported cases $E$, while the latter one says that the effective reproduction number $\Rt$ is assumed to be a piece-wise constant function with jump every one day time interval. 
	
	Discretizing the model using the forward Euler method, we obtain the following discrete-time augmented SIRD model:
	\begin{eqnarray}
		S(k+1) &=& \left(1-\frac{(\gamma+\kappa)\Delta t}{N}\Rt(k)I(k)\right)S(k),\label{S}\\
		I(k+1) &=& (1-(\gamma+\kappa)\Delta t)I(k)+\frac{(\gamma+\kappa)\Delta t}{N}\Rt(k)I(k)S(k),\\
		R(k+1) &=& R(k)+\gamma\Delta t I(k),\\
		D(k+1) &=& D(k)+\kappa\Delta t I(k),\\
		E(k+1) &=& (\gamma+\kappa)\Delta t I(k)+(1-\Delta t)E(k),\\
		\Rt(k+1) &=& \Rt(k).\label{Rt}
	\end{eqnarray}
	
	Our method computes a new estimate of $\Rt$ based on new reported cases. Since their frequency is low (could be once a day), the reported data can be interpolated using, e.g., a modified Akima cubic Hermite interpolation, such that it fits with the time step $\Delta t$. In our simulation, the time step $\Delta t$ is chosen as 0.01, i.e., 100 time discretization within one day interval. The confidence interval of our estimated $\Rt$ is determined by computing the reproduction number for different values of the infectious period $T_i$ within a certain interval.
	
	To simplify the presentation, we define the augmented state vector 
	\begin{eqnarray}
		\bm{x}(k+1) &=&\begin{pmatrix}S(k+1)\\I(k+1)\\R(k+1)\\D(k+1)\\E(k+1)\\\Rt(k+1)\end{pmatrix}, 
	\end{eqnarray}
	and as such, the discrete-time augmented SIRD model \eqref{S}-\eqref{Rt} can be written as follows
	\begin{eqnarray}
		\bm{x}(k+1) &=& \bm{f}(\bm{x}(k))+\bm{w}(k),\label{non}
	\end{eqnarray}
	where $\bm{f}$ is the nonlinear term written in the right hand side of \eqref{S}-\eqref{Rt} and $\bm{w}$ is introduced as an uncertainty to model the inaccuracies due to simplification in the modelling. The uncertainty is assumed to be a zero mean Gaussian white noise with known covariance $\bm{Q}_F$. This is to simplify the calculation since the actual epidemic data usually follow Gamma distribution. In practice, $\bm{Q}_F$ can be considered as a tuning parameter for the EKF. Thus, the transmission model becomes a discrete-time stochastic augmented SIRD model. 
	
	Reported cases, such as the number of active cases and the cumulative numbers of recovered and death, can be incorporated into the model using the following output vector
	\begin{eqnarray}
		\bm{y}(k+1) &=& \bm{C}\bm{x}(k)+\bm{v}(k).
	\end{eqnarray}
	Here, $\bm{v}$ denotes uncertainties due to false testing results. We also assume the uncertainty to be a zero mean Gaussian white noise with known covariance $\bm{R}_F$. As well as $\bm{Q}_F$, $\bm{R}_F$ can also be considered as a tuning parameter. Following the available data that include $I$, $R$, $D$, and $E$, the data/measurement matrix $\bm{C}$ is taken to be
	\begin{eqnarray}
		\bm{C} = \begin{pmatrix}1 & 0 & 0 & 0 & 0 & 0\\0 & 1 & 0 & 0 & 0 & 0\\0 & 0 & 1 & 0 & 0 & 0\\0 & 0 & 0 & 1 & 0 & 0\\0 & 0 & 0 & 0 & 1 & 0\end{pmatrix}.
	\end{eqnarray}
	
	\section*{A Two-Stage Filtering Method}
	
	A two-stage filtering method is used to estimate the daily reproduction number $\Rt$. The method consists of the EKF and a low-pass filter. In the first stage of estimation, the EKF is used to estimate the state variables and the value of $\Rt$ under uncertainties in the number of reported cases. Afterwards, the low pass filter is used to remove short term fluctuations of the reported cases that can be caused by delays in the reporting. %A short term fluctuation can be caused when an accumulation of cases from previous days is reported in one day. 
	For example, suddenly in Denmark there were 893 recovered patients reported on 1 April 2020, in contrast to the previous days from 16 February 2020 onwards when there was no recovery reported at all. Such an accumulated delay can cause a falsely decreasing value of $\Rt$.
	
	The EKF is an extension of Kalman filter for nonlinear systems. The Kalman filter itself is based on a recursive Bayesian estimation and is an optimal linear filter. The idea of EKF is to linearize the non-linearity around its estimate. Due to that linearization, the optimality and stability of the EKF cannot be guaranteed. However, if the non-linearity is not severe, the EKF can give a reasonably good estimate. 
	
	Let us denote $\bm{\hat{x}}(k)$ as an estimated vector state from the EKF. Applying first-order Taylor series expansion to $\bm{f}$ at  $\bm{\hat{x}}(k)$, we obtain
	\begin{eqnarray}
		\bm{f}(\bm{x}(k)) = \bm{f}(\bm{\hat{x}}(k)) + \bm{J}_{\bm{f}}(\bm{\hat{x}}(k))(\bm{x}(k)-\bm{\hat{x}}(k)),
	\end{eqnarray}
	where $\bm{J}_{\bm{f}}(\bm{\hat{x}}(k))$ is the Jacobian matrix of $\bm{f}$, given by:
	\begin{eqnarray}
		\bm{J}_{\bm{f}}(\bm{\hat{x}}(k)) =
		\begin{pmatrix}
			J_{11}(\bm{\hat{x}}(k)) & J_{12}(\bm{\hat{x}}(k)) & 0 & 0 & 0 & J_{16}(\bm{\hat{x}}(k))\\
			J_{21}(\bm{\hat{x}}(k)) & J_{22}(\bm{\hat{x}}(k)) & 0 & 0 & 0 & J_{26}(\bm{\hat{x}}(k))\\
			0 & \gamma\Delta t & 1 & 0 & 0 & 0\\
			0 & \kappa\Delta t & 0 & 1 & 0 & 0\\
			0 & (\gamma+\kappa)\Delta t & 0 & 0 & 1-\Delta t & 0\\
			0 & 0 & 0 & 0 & 0 &  1
		\end{pmatrix},
	\end{eqnarray}
	where
	\begin{eqnarray}
		J_{11}(\bm{\hat{x}}(k)) &=& 1-\frac{(\gamma+\kappa)\Delta t}{N}\hat{\Rt}(k)\hat{I}(k),\\
		J_{12}(\bm{\hat{x}}(k)) &=& -\frac{(\gamma+\kappa)\Delta t}{N}\hat{\Rt}(k)\hat{S}(k),\\
		J_{16}(\bm{\hat{x}}(k)) &=& -\frac{(\gamma+\kappa)\Delta t}{N}\hat{I}(k)\hat{S}(k),\\
		J_{21}(\bm{\hat{x}}(k)) &=& \frac{(\gamma+\kappa)\Delta t}{N}\hat{\Rt}(k)\hat{I}(k),\\
		J_{22}(\bm{\hat{x}}(k)) &=& 1-(\gamma+\kappa)\Delta t+\frac{(\gamma+\kappa)\Delta t}{N}\hat{\Rt}(k)\hat{S}(k),\\
		J_{26}(\bm{\hat{x}}(k)) &=& \frac{(\gamma+\kappa)\Delta t}{N}\hat{I}(k)\hat{S}(k).
	\end{eqnarray}
	
	The EKF consists of two steps: predict and update. The discrete-time stochastic augmented SIRD model is used to predict the next state and covariance and update them after obtaining new data/measurement. The EKF can be considered as one of the simplest dynamic Bayesian networks. While the EKF calculates estimates of the true values of states recursively over time using incoming measurements and a mathematical process model, recursive Bayesian estimation calculates estimates of an unknown probability density function recursively over time using incoming measurements and a mathematical process model \cite{martin77}. Let $\bm{\hat{x}}(n|m)$ denotes the estimate of $\bm{x}$ at time $n$ given observations up to and including at time $m\leq n$. The Kalman filter algorithm is given as follows \cite{simon06}
	
	\textbf{Predict}
	\begin{eqnarray}
		\bm{\hat{x}}(k+1|k) &=& \bm{f}(\bm{\hat{x}}(k|k))\\
		\bm{P}(k+1|k) &=& \bm{J}_{\bm{f}}(\bm{\hat{x}}(k|k)) \bm{P}(k|k) \bm{J}_{\bm{f}}(\bm{\hat{x}}(k|k))^\intercal + \bm{Q}_F(k)
	\end{eqnarray}
	
	\textbf{Update}
	\begin{eqnarray}
		\bm{\tilde{y}}(k+1) &=& \bm{y}(k+1)-\bm{C}\bm{\hat{x}}(k+1|k)\\
		\bm{K}(k+1) &=& \bm{P}(k+1|k)\bm{C}^\intercal\left(\bm{C}\bm{P}(k+1|k)\bm{C}^\intercal+\bm{R}_F(k)\right)^{-1}\\
		\bm{\hat{x}}(k+1|k+1) &=& \bm{\hat{x}}(k+1|k) + \bm{K}(k+1) \bm{\tilde{y}}(k+1)\\
		\bm{P}(k+1|k+1) &=& \left(\bm{I}-\bm{K}(k+1)\bm{C}\right)\bm{P}(k+1|k)
	\end{eqnarray}
	
	Here $\bm{P}(k|k)$ denotes a posteriori estimate covariance matrix. In the second stage, a low pass filter based on a rational transfer function is used to remove short term fluctuation at time step $k$, and is given by
	\begin{eqnarray}
		\bm{\hat{y}}(k) = \frac{1}{y_n}\left(\bm{\hat{x}}(k)+\bm{\hat{x}}(k-1)+\cdots+\bm{\hat{x}}(k-y_n+1)\right),
	\end{eqnarray}
	where $y_n$ is a window length along the data. In our case, we choose $y_n=\frac{3}{\Delta t}$.
	
	To evaluate the quality of the estimate, we calculate a Relative Root Mean Square Error (RRMSE) between the estimated and reported cases. The RRMSE is defined as
	\begin{eqnarray}
		\text{RRMSE} = \frac{1}{N_d}\sum_{i=1}^{N_d} \frac{\|X_i-\hat{X}_i\|^2}{\|X_i\|^2},
	\end{eqnarray}
	where $N_d$ is the number of observed days and $X\in\{I,D,R,E\}$.
	
	\section*{Case study: Scandinavian countries}
	
	\begin{figure}[tbhp]
		\begin{center}
			% include first image
			\subfloat[Denmark]{\includegraphics[width=0.95\linewidth]{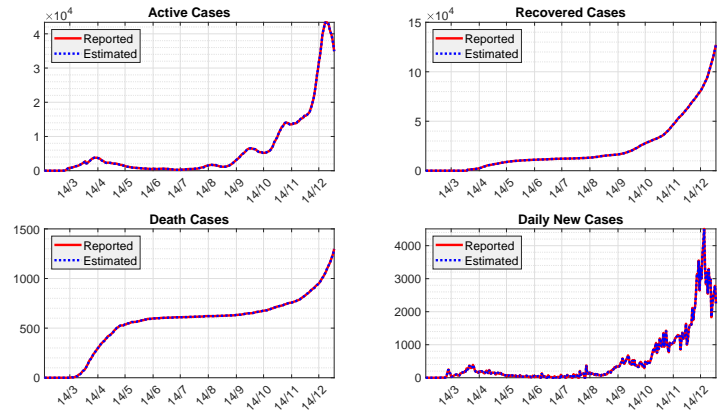}}\\
			\subfloat[Sweden]{\includegraphics[width=0.95\linewidth]{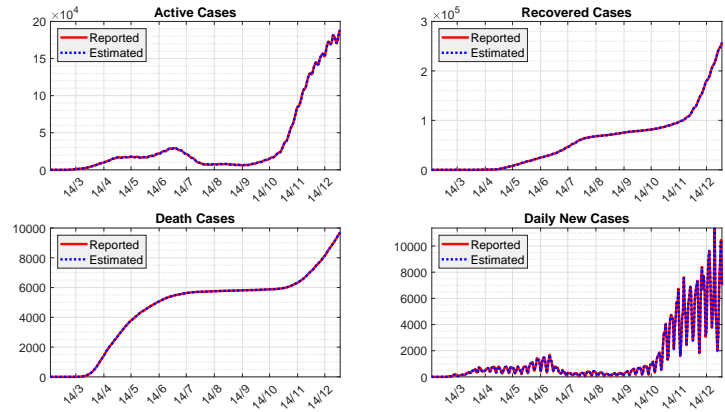}}  
		\end{center}
		\caption{(Continued)}
		\label{fig1a}
	\end{figure}
\setcounter{figure}{0}
	\begin{figure}[tbhp]
%		\ContinuedFloat	
		\begin{center}
			% include first image
			\subfloat[Norway]{\includegraphics[width=0.95\linewidth]{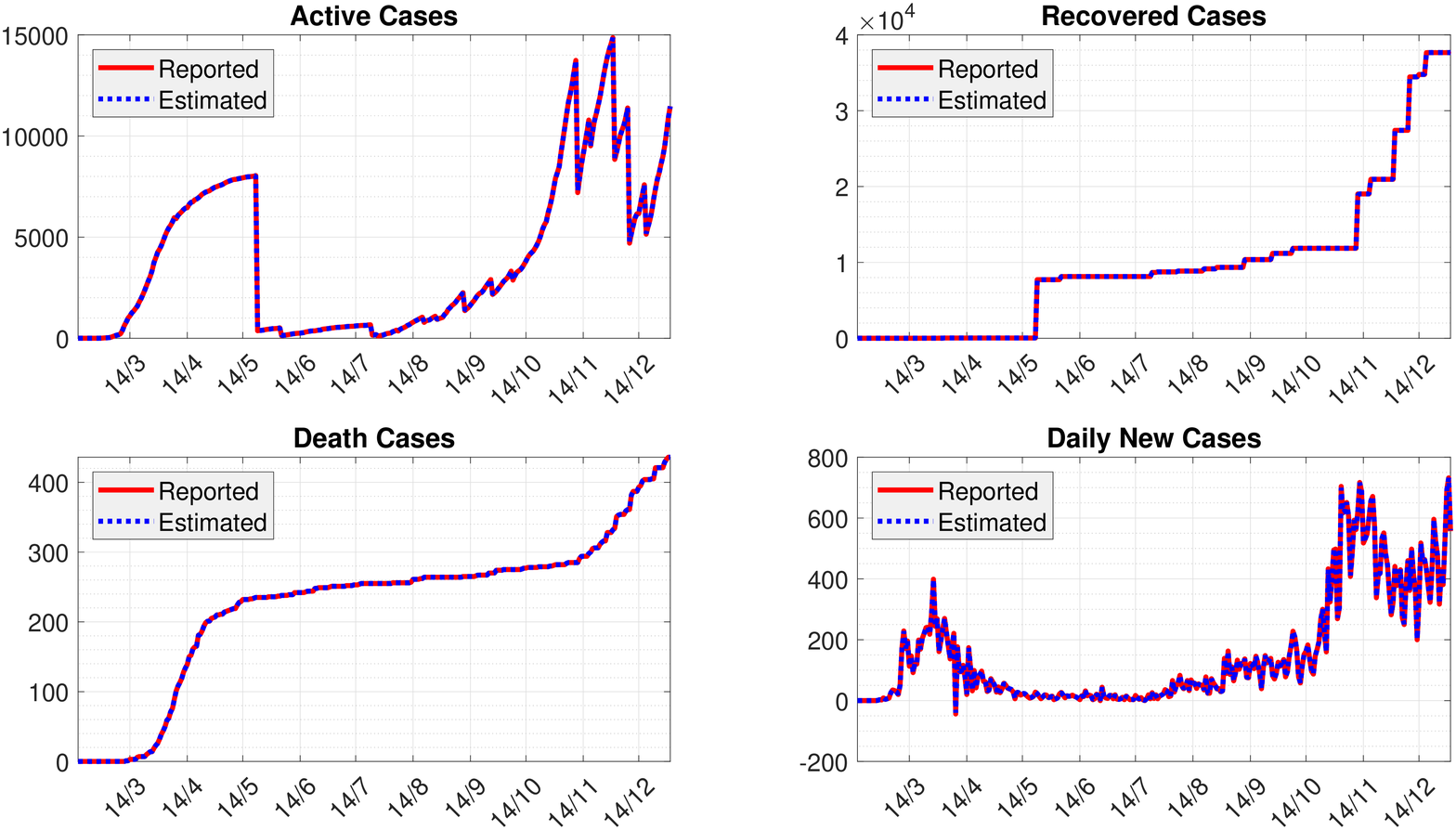}}
		\end{center}
		\caption{{\color{blue}Comparison between reported and estimated cases for active (I), recovered (R), death (D), and daily new cases (E) from the three Scandinavian countries.}}
		\label{fig1}
	\end{figure}
	
	In this section, we apply our method to study viral transmission of COVID-19 in Denmark, Sweden, and Norway. All datasets and $\mathsf{MATLAB}$ code are available on GitHub (\url{https://github.com/agusisma/covid19}). As of January 2021, the three Scandinavian countries have higher cumulative testing rate compared to other parts of the world, with Denmark held a record with 260 tests per 1000 population. During that time the daily test-positivity rate is below 1\% for Denmark and Norway, while Sweden is at 2.9\% \cite{Erica}. These numbers are good indications about the testing capacity in the Scandinavian countries and may describe the dynamics of the transmission better with respect to asymptomatic cases. The countries also have a different approach in their public measures in responding to COVID-19, e.g., Sweden did not implement a strict lockdown, unlike its Nordic neighbouring countries. 
	
	We plot the observed incidence of COVID-19 in Denmark, Sweden, and Norway in Fig.\ \ref{fig1}. We also plot in the same figure estimated numbers computed using our method, where good agreement is obtained. For all estimation, the process and observation covariance matrices are considered as tuning parameters and are chosen as $\bm{Q}_F=\text{diag}{(10\;10\;10\;10\;5\;0.2)}$ and $\bm{R}_F=\text{diag}{(100\;10\;10\;5\;1)}$, respectively. These parameters are obtained from trial and error and are chosen such that the RRMSE between the estimated and reported data are sufficiently small. In our case study, the RRMSE are shown in Table \ref{table:1}. Here, we can observe the method provides relatively small estimation errors for all countries. Norway has the largest error, which can be attributed to the lack of daily update of the active and recovered cases. In our simulation, we use the same tuning parameters. The error can be reduced by using different value of $\bm{Q}_F$ and $\bm{R}_F$.
	
	\begin{table}[h!]
		\centering
		\begin{tabular}{lcccc|c}
			\hline\hline
			\multicolumn{6}{c}{RRMSE} \\
			\cline{2-6}
			Country  & I & R & D & E & Total\\
			\hline\hline
			Denmark  & 7.8943e-05 & 0.1396  & 1.1640e-04  &  6.0490e-05 & 0.1399    \\
			Sweden   & 4.9208e-05 &  0.0155 & 1.3375e-04  &  0.0102  &  0.0259 \\
			Norway   & 0.0011 &  0.0682 & 7.9477e-05  & 0.1631   &  0.2326  \\
			\hline\hline
		\end{tabular}
		\caption{RRMSE of the two-stage filtering method for the three Scandinavian countries.}
		\label{table:1}
	\end{table}
	
	In applying our method, we also compare it with two commonly used methods to estimate transmission parameters, namely the sequential Bayesian method of Bettencourt and Ribeiro \cite{bett08} that provides an approximation of the basic reproduction number, and the instantaneous method by Fraser \cite{fras07} that is implemented with a Bayesian analysis \cite{nish09,cori13}. The former method exploits the new reported incidence, while the latter one uses the distribution of the serial interval.
	
	\begin{figure}[tbhp]
		\centering
		% include first image
		\subfloat[Denmark]{\includegraphics[width=0.95\linewidth]{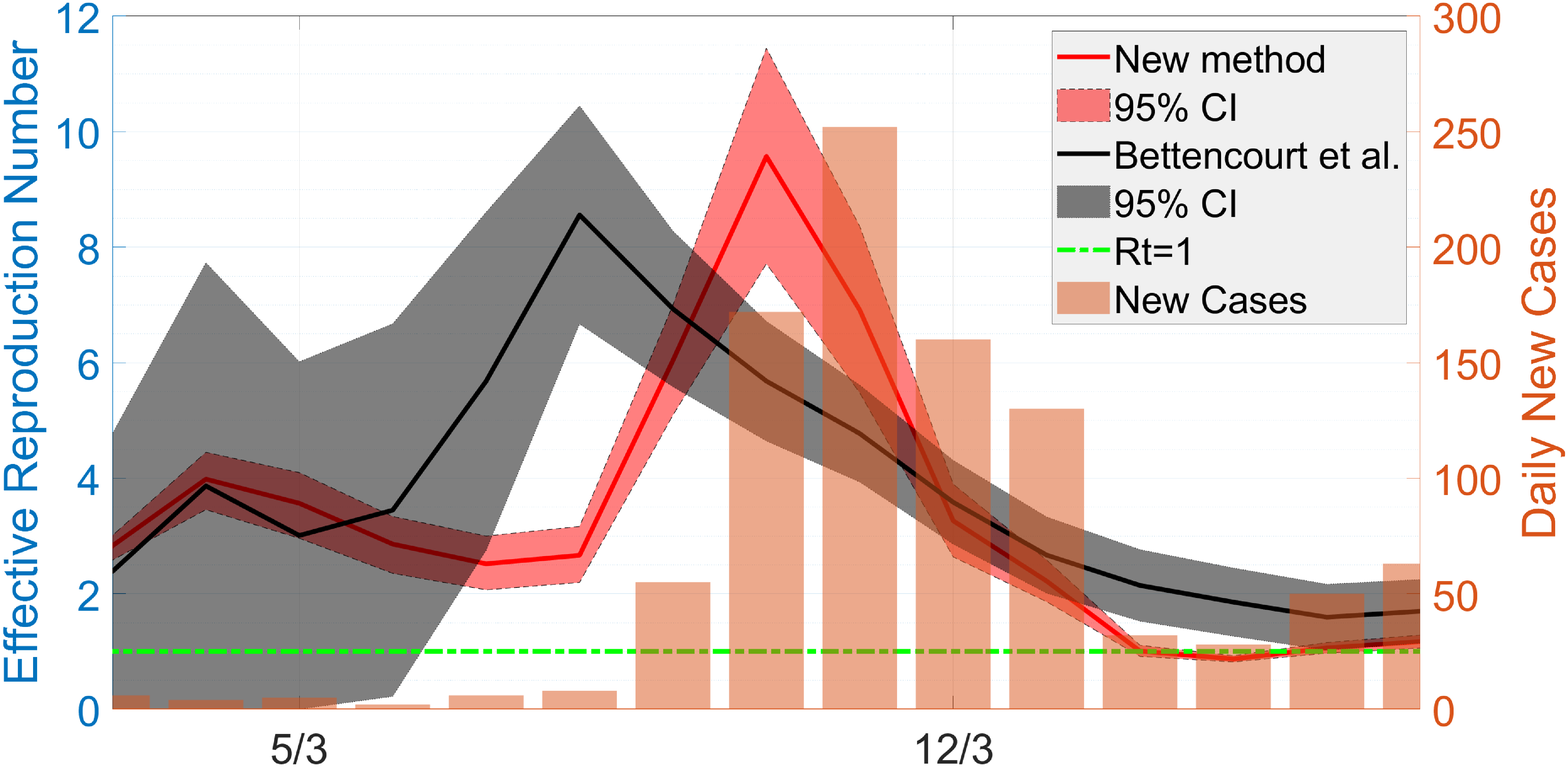}}\\
		\subfloat[Sweden]{\includegraphics[width=0.95\linewidth]{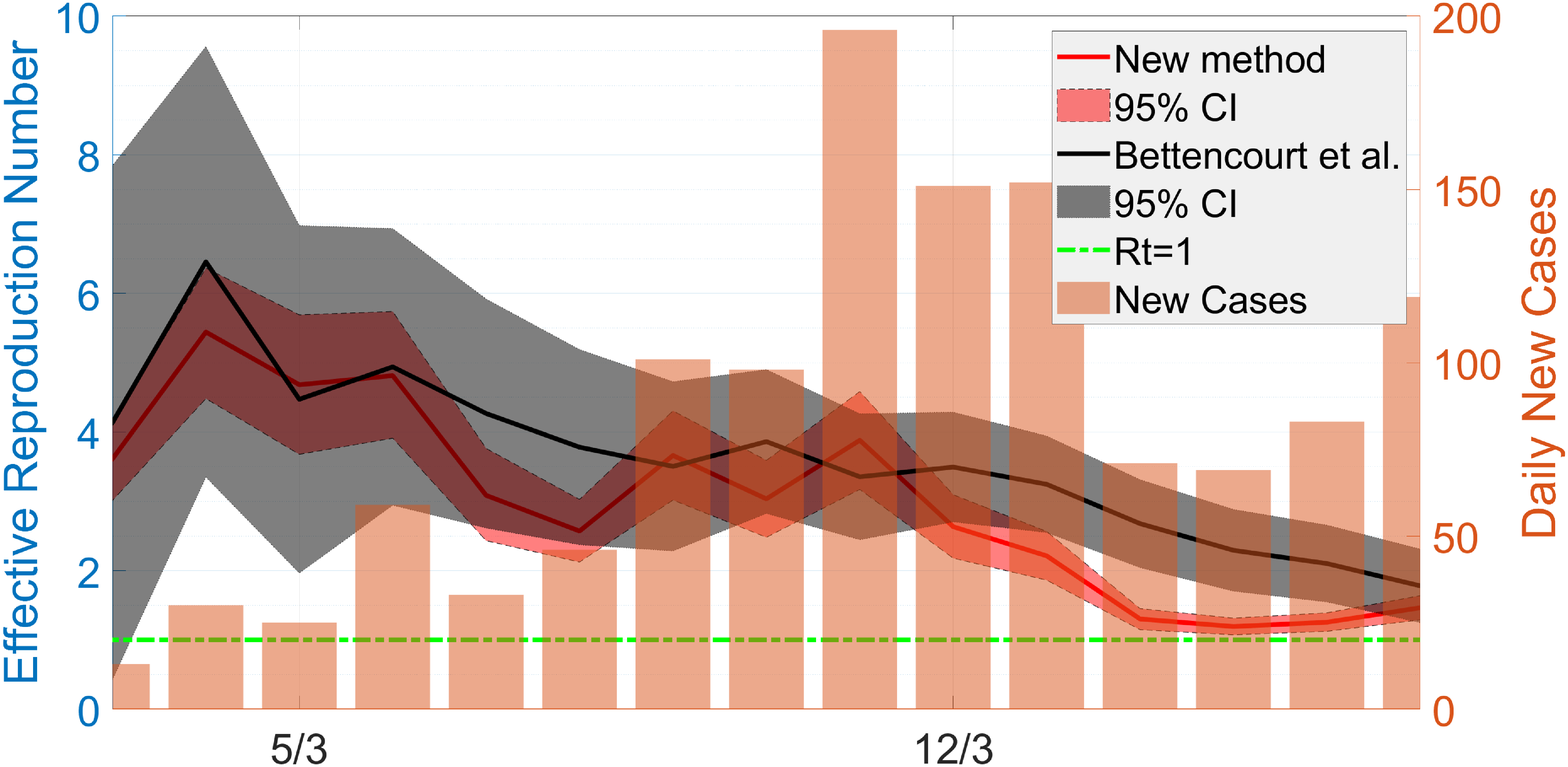}}
		\caption{(Continued)}
		\label{fig2a}
	\end{figure}
\setcounter{figure}{3}
	\begin{figure}[tbhp]
		\ContinuedFloat	\centering
		% include first image
		\subfloat[Norway]{\includegraphics[width=0.95\linewidth]{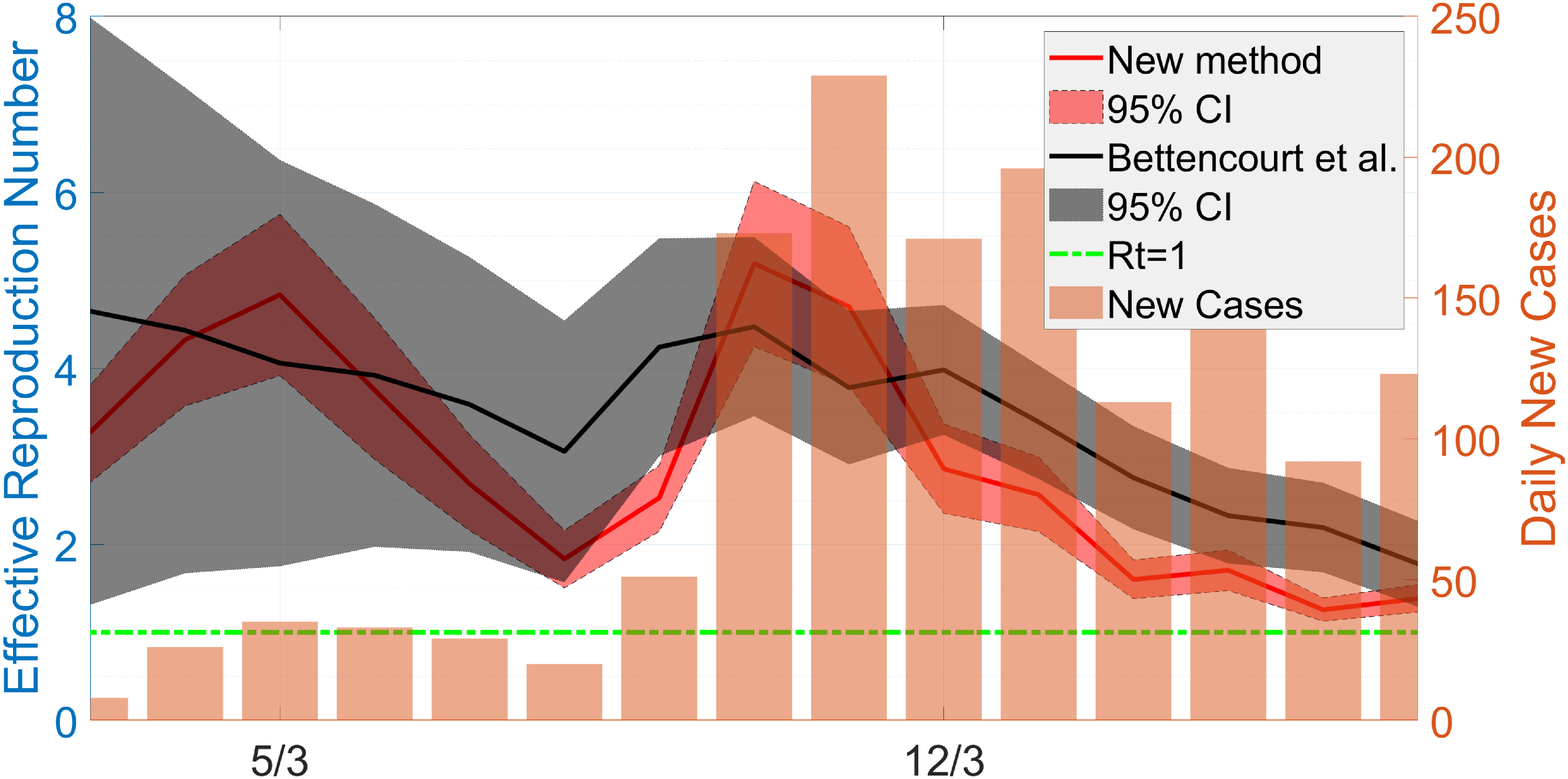}}
		\caption{{\color{blue}Comparison of the estimated reproduction number at the early stage of the pandemic between our proposed method and Bettencourt and Ribeiro \cite{bett08}.}}
		\label{fig2}
	\end{figure}
	
	\begin{table}[h!]
		\centering
		\begin{tabular}{||l c c||} 
			\hline
			& Current method & Bettencourt \& Ribeiro \cite{bett08} \\ [0.5ex] 
			\hline\hline
			Denmark & 9.6 [95\% CI: 7.7-11.4] & 8.6 [95\% CI: 6.7-10.5]  \\ 
			Sweden & 5.4 [95\% CI: 4.9-6.4] & 6.5 [95\% CI: 3.3-9.6] \\ 
			Norway & 5.2 [95\% CI: 4.2-6.1] & 4.6 [95\% CI: 1.3-7.9] \\ 
			[1ex]
			\hline
		\end{tabular}
		\caption{Estimation of the basic reproduction number $\Rn$ using our method and Bettencourt and Ribeiro \cite{bett08}.}
		\label{table:2}
	\end{table}
	
	First, we compare our method with Bettencourt and Ribeiro \cite{bett08}, that allows sequential
	estimation of the basic reproduction number at the initial stage when the growth is still exponential. While the two methods are based on the SIR model, Bettencourt and Ribeiro \cite{bett08} use new incidence data and the result is filtered using a five-day moving average filter. In Fig.\ \ref{fig2}, we plot the comparison and summarise the basic reproduction numbers that are taken to be the maximum of the curves in Table \ref{table:2}. It is interesting to note how the methods give rather similar estimations. %, except in the case of Norway. 
	This indicates that our method gives comparable results to those of \cite{bett08}.
	
	\begin{figure}[tbhp]
		\centering
		% include first image
		\subfloat[Denmark]{\includegraphics[width=0.95\linewidth]{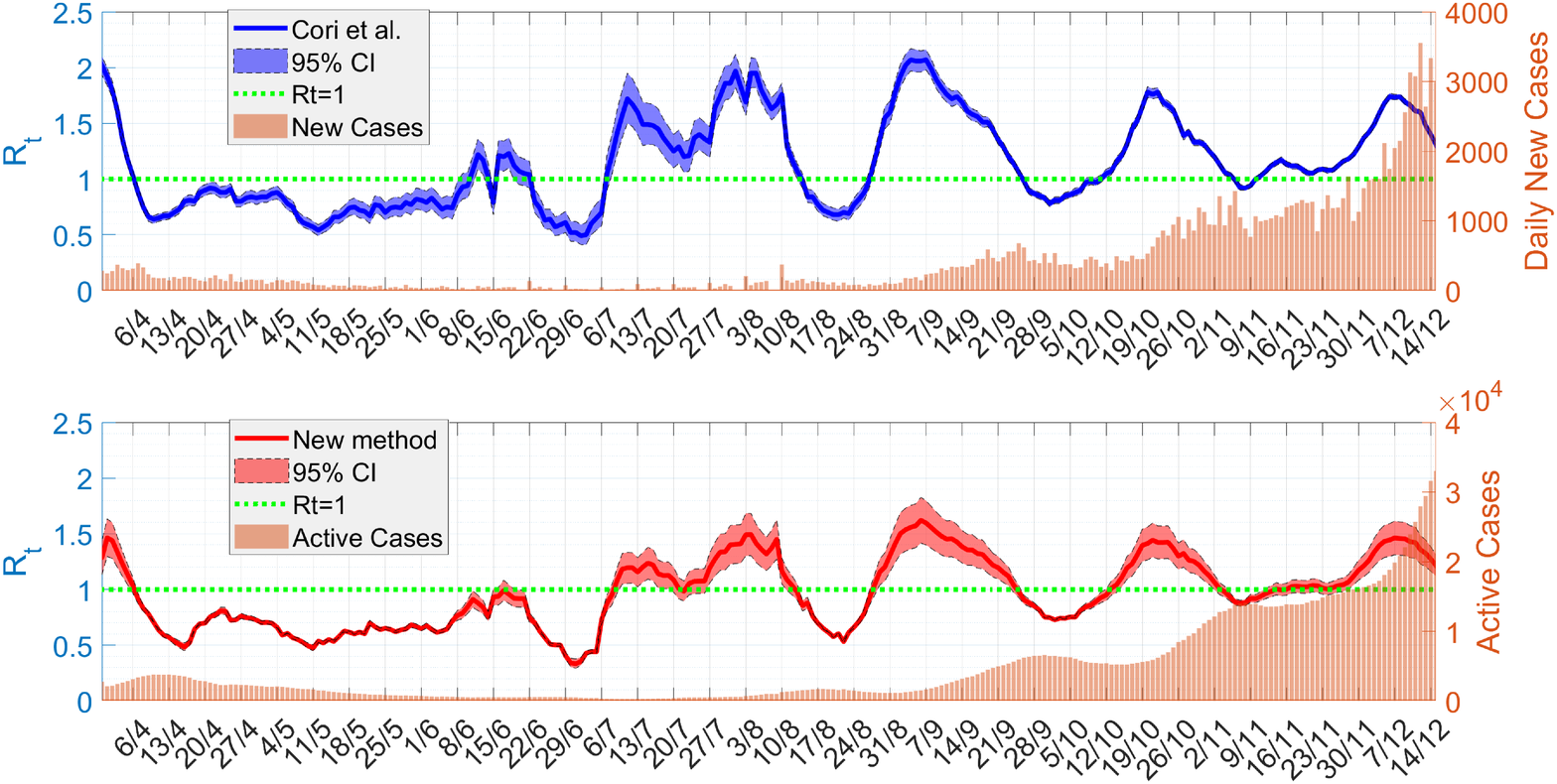}}\\
		\subfloat[Sweden]{\includegraphics[width=0.95\linewidth]{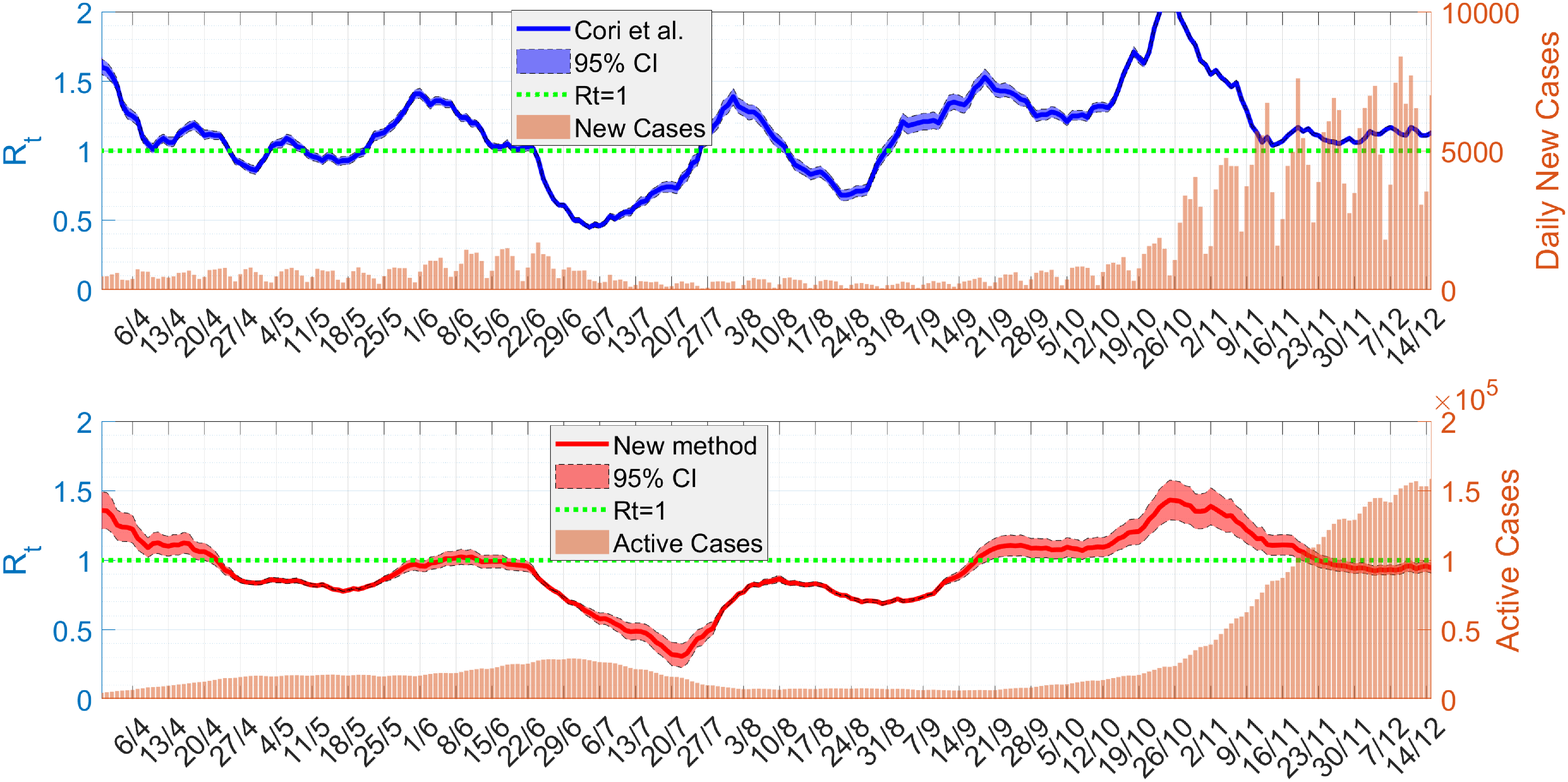}}
		\caption{(Continued)}
		\label{fig3a}
	\end{figure}
\setcounter{figure}{4}
	\begin{figure}[tbhp]
		\ContinuedFloat\centering
		\subfloat[Norway]{\includegraphics[width=0.95\linewidth]{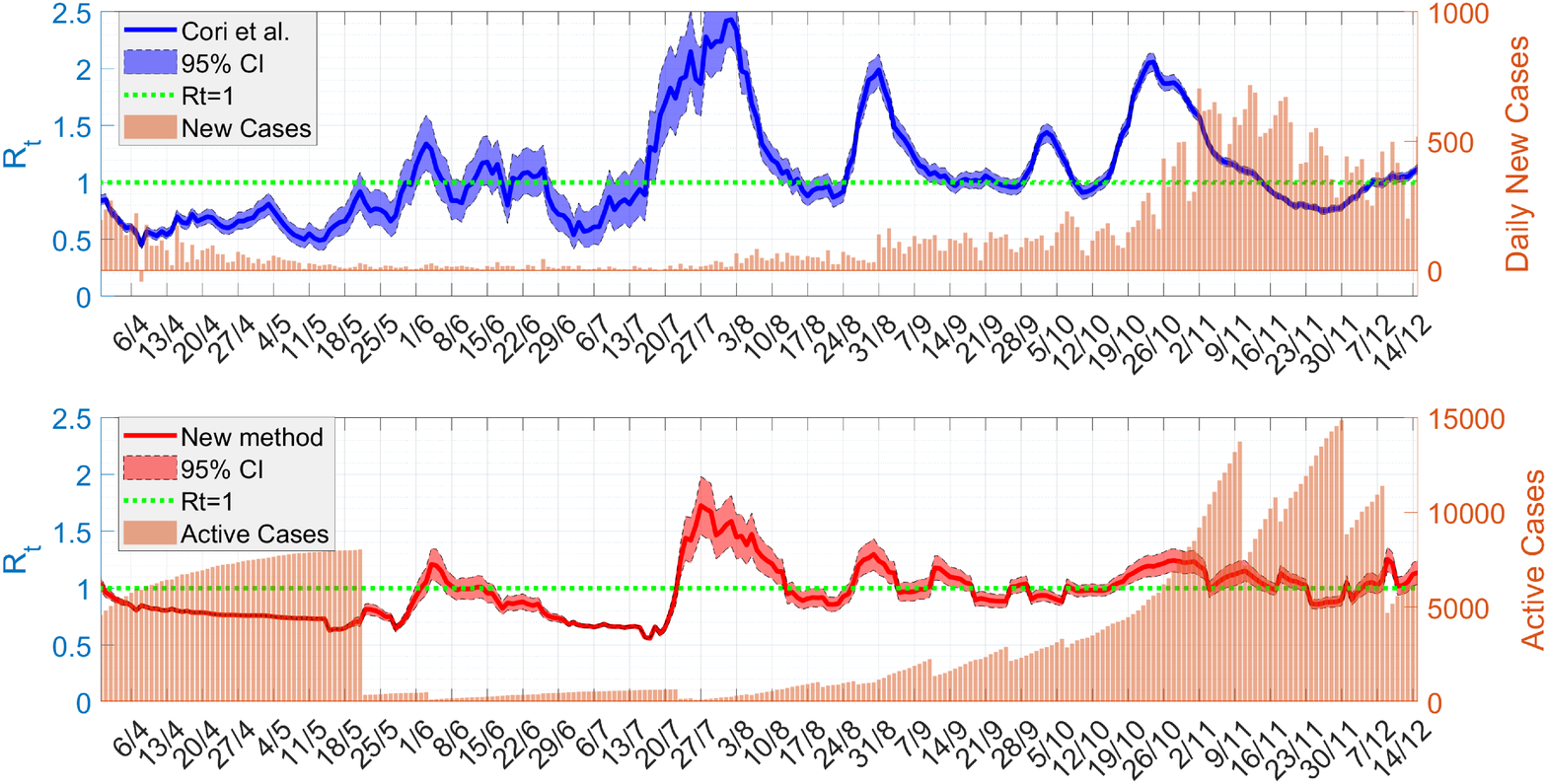}}
		\caption{{\color{blue}Comparison of the estimated time-varying effective reproduction number between our method and Cori, et al \cite{cori13}.}}
		\label{fig3}
	\end{figure}
	
	Finally, we plot the time-varying effective reproduction number in Fig.\ \ref{fig3}. Here, we compare our results with those using Cori et al.\ \cite{cori13}. The method of \cite{cori13} utilises the disease serial interval, which we approximate using a shifted Gamma distribution \cite{cori13} with mean 4.7 and standard deviation 2.9 \cite{nish20}. The prior belief for the value of $\Rt$ is taken to be Gamma function with mean and standard deviation 5. We do not average out the data of daily new cases, but instead take the likelihood estimation of a new case at one day to depend also on the estimation of the previous three days. 
	
	In Fig.\ \ref{fig3} we obtain that the two methods give the plot of $\Rt$ with the same trend, indicating that our method is also comparable with \cite{cori13}. There is a delay of about four days in the trend, especially with the time when the reproduction number curve crossed the horizontal axis. %, compared to that of \cite{cori13}.
	The delay is caused by the peaks of new daily cases and active ones that also differ by about the same days. 
	
	A different trend especially at later times between the methods appears in Fig.\ \ref{fig3}(c) for Norway. The curve from our method is quite smooth, while it is rather fluctuating in that using Cori et al.\ \cite{cori13}. The discrepancy is caused by the active and recovered cases that apparently were not updated regularly, in contrast to the new positive cases needed by the method of \cite{cori13}. The unreported recovery cases were all released at once on May 22, 2020, see Fig.\ \ref{fig1}(c).
	
	\section*{Conclusion and Future Work}
	
	Many mathematical models and estimation methods have been developed to estimate several types of reproduction numbers during epidemic outbreaks. Here, we provide a novel method exploiting reported active, recovered and death cases using the SIR model as an underlying approach. This new method offers several advantages compared to existing methods: (i) from modeling point of view, the resulting $\Rt$ value can follow the dynamics of the model suggested, so it is possible to develop it further if the model chosen has a higher complexity, (ii) the estimation method can still be expanded in terms of statistical view, and (iii) the method does not need information about serial intervals. In the case that the data provided in time series do not change much or instead have drastic changes, such as accumulating at a certain time, the resulting $\Rt$ value will show the same spikes and serrations. As a result, the latest information from data dynamics can be more elaborated.
	
	By applying the method to COVID-19 cases in the Scandinavian countries and comparing the results to commonly used methods due to \cite{bett08} and \cite{cori13}, we showed that our model is comparable, which expectedly will allow for fast assessment of the reproduction number in new outbreaks. Using the method to forecast and critically assess incidence data in countries with high under-reporting, such as Indonesia, is addressed for future work.
	
	\section*{Data and Code Availability}
	
	Data and codes can be found here: \url{https://github.com/agusisma/covid19}
	
%	\bibliography{sample}

\begin{thebibliography}{10}
		\expandafter\ifx\csname url\endcsname\relax
		\def\url#1{\texttt{#1}}\fi
		\expandafter\ifx\csname urlprefix\endcsname\relax\def\urlprefix{URL }\fi
		\expandafter\ifx\csname href\endcsname\relax
		\def\href#1#2{#2} \def\path#1{#1}\fi
		
		\bibitem{ncov20}
		{The 2019-nCoV Outbreak Joint Field Epidemiology Investigation Team}, An
		outbreak of {NCIP} (2019-{nCoV}) infection in {C}hina - {W}uhan, {H}ubei
		province, China CDC Weekly 2~(5) (2020) 79--80.
		\newblock \href {http://dx.doi.org/10.46234/ccdcw2020.022}
		{\path{doi:10.46234/ccdcw2020.022}}.
		
		\bibitem{yliu20}
		Y.~Liu, A.~Gayle, A.~Wilder-Smith, J.~Rocklov, {The reproductive number of
			COVID-19 is higher compared to SARS Coronavirus}, Journal of Travel Medicine
		27~(2) (2020) 1--4.
		\newblock \href {http://dx.doi.org/10.1093/jtm/taaa021}
		{\path{doi:10.1093/jtm/taaa021}}.
		
		\bibitem{dela19}
		P.~Delamater, E.~Street, T.~Leslie, Y.~Yang, K.~Jacobsen, {Complexity of the
			basic reproduction number ($R_0$)}, Emerging Infectious Diseases 25~(1)
		(2019) 1--4.
		\newblock \href {http://dx.doi.org/10.3201/eid2501.171901}
		{\path{doi:10.3201/eid2501.171901}}.
		
		\bibitem{sanc20}
		S.~Sanche, Y.~Lin, C.~Xu, E.~Romero-Severson, N.~Hengartner, R.~Ke, {High
			contagiousness and rapid spread of severe acute respiratory syndrome
			Coronavirus 2}, Emerging Infectious Diseases 26~(7) (2020) 1--8.
		\newblock \href {http://dx.doi.org/10.3201/eid2607.200282}
		{\path{doi:10.3201/eid2607.200282}}.
		
		\bibitem{fras07}
		C.~Fraser, Estimating individual and household reproduction numbers in an
		emerging epidemic, PLoS One 2~(1) (2007) 1--12.
		\newblock \href {http://dx.doi.org/10.1371/journal.pone.0000758}
		{\path{doi:10.1371/journal.pone.0000758}}.
		
		\bibitem{wall04}
		J.~Wallinga, P.~Teunis, Different epidemic curves for severe acute respiratory
		syndrome reveal similar impacts of control measures, American Journal of
		Epidemiology 160~(6) (2004) 509--516.
		\newblock \href {http://dx.doi.org/10.1093/aje/kwh255}
		{\path{doi:10.1093/aje/kwh255}}.
		
		\bibitem{cauc06}
		S.~Cauchemez, P.-Y. Boelle, C.~Donnelly, N.~Ferguson, G.~Thomas, G.~Leung,
		A.~Hedley, R.~Anderson, A.-J. Valleron, Real-time estimates in early
		detection of {SARS}, Emerging Infectious Diseases 12~(1) (2006) 1--4.
		\newblock \href {http://dx.doi.org/10.3201/eid1201.050593}
		{\path{doi:10.3201/eid1201.050593}}.
		
		\bibitem{whit08}
		L.~White, M.~Pagano, Transmissibility of the {I}nfluenza virus in the 1918
		pandemic, PLoS One 3~(1) (2008) 1--6.
		\newblock \href {http://dx.doi.org/10.1371/journal.pone.0001498}
		{\path{doi:10.1371/journal.pone.0001498}}.
		
		\bibitem{nish10}
		H.~Nishiura, G.~Chowell, M.~Safan, C.~Castillo-Chavez, Pros and cons of
		estimating the reproduction number from early epidemic growth rate of
		{I}nfluenza a ({H1N1}) 2009, Theoretical Biology and Medical Modelling 7~(1)
		(2010) 1--13.
		\newblock \href {http://dx.doi.org/10.1186/1742-4682-7-1}
		{\path{doi:10.1186/1742-4682-7-1}}.
		
		\bibitem{bernard18}
		B.~Cazelles, C.~Champagne, J.~Dureau, {Accounting for non-stationarity in
			epidemiology by embedding time-varying parameters in stochastic models}, PLOS
		Computational Biology 15~(5) (2018) e1007062.
		\newblock \href {http://dx.doi.org/10.1371/journal.pcbi.1006211}
		{\path{doi:10.1371/journal.pcbi.1006211}}.
		
		\bibitem{bett08}
		L.~Bettencourt, R.~Ribeiro, {Real time Bayesian estimation of the epidemic
			potential of emerging infectious diseases}, PLoS One 3~(5) (2008) 1--9.
		\newblock \href {http://dx.doi.org/10.1371/journal.pone.0002185}
		{\path{doi:10.1371/journal.pone.0002185}}.
		
		\bibitem{Aaron}
		H.~Zhao, X.~Lu, Y.~Deng, Y.~Tang, J.~Lu, {COVID-19: asymptomatic carrier
			transmission is an underestimated problem}, Epidemiology \& Infection 148~(1)
		(2020) 1--3.
		\newblock \href {http://dx.doi.org/10.1017/S0950268820001235}
		{\path{doi:10.1017/S0950268820001235}}.
		
		\bibitem{cori13}
		A.~Cori, N.~Ferguson, C.~Fraser, S.~Cauchemez, A new framework and software to
		estimate time-varying reproduction numbers during epidemics, American Journal
		of Epidemiology 178~(9) (2013) 1505--1512.
		\newblock \href {http://dx.doi.org/10.1093/aje/kwt133}
		{\path{doi:10.1093/aje/kwt133}}.
		
		\bibitem{Francisco}
		F.~Arroyo-Marioli, F.~Bullano, S.~Kucinskas, C.~Rondon-Moreno, {Tracking R of
			COVID-19: A new real-time estimation using the Kalman filter}, PLOS One
		16~(1) (2021) 1--16.
		\newblock \href {http://dx.doi.org/10.1371/journal.pone.0244474}
		{\path{doi:10.1371/journal.pone.0244474}}.
		
		\bibitem{Lan20}
		K.~Kai-Wang~To, O.~Tak-Yin~Tsang, W.-S. Leung, A.~Raymond~Tam, T.-C. Wu,
		D.~Christopher~Lung, C.-Y. Cyril~Yip, J.-P. Cai, J.~Man-Chun~Chan,
		T.~Shiu-Hong~Chik, D.~Pui-Ling~Lau, C.~Yau-Chung~Choi, L.-L. Chen, W.-M.
		Chan, K.-H. Chan, J.~Daniel~Ip, A.~Chin-Ki~Ng, R.~Wing-Shan~Poon, C.-T. Luo,
		V.~Chi-Chung~Cheng, J.~Fuk-Woo~Chan, I.~Fan-Ngai~Hung, Z.~Chen, H.~Chen,
		K.-Y. Yuen, Temporal profiles of viral load in posterior oropharyngeal saliva
		samples and serum antibody responses during infection by {SARS-CoV-2}: an
		observational cohort study, The Lancet Infectious Diseases 20~(1) (2020)
		565--574.
		\newblock \href {http://dx.doi.org/10.1016/S1473-3099(20)30196-1}
		{\path{doi:10.1016/S1473-3099(20)30196-1}}.
		
		\bibitem{martin77}
		C.~Masreliez, R.~Martin, Robust bayesian estimation for the linear model and
		robustifying the {K}alman filter, IEEE Transactions on Automatic Control
		22~(3) (1977) 361--371.
		\newblock \href {http://dx.doi.org/10.1109/TAC.1977.1101538}
		{\path{doi:10.1109/TAC.1977.1101538}}.
		
		\bibitem{simon06}
		D.~Simon, Optimal State Estimation: Kalman, H Infinity, and Nonlinear
		Approaches, 1st Edition, Wiley-Interscience, 2006.
		
		\bibitem{Erica}
		E.~A. Yarmol-Matusiak, L.~E. Cipriano, S.~Stranges, {A comparison of COVID-19
			epidemiological indicators in Sweden, Norway, Denmark, and Finland},
		Scandinavian Journal of Public Health 49~(1) (2021) 69--78.
		\newblock \href {http://dx.doi.org/doi.org/10.1177/1403494820980264}
		{\path{doi:doi.org/10.1177/1403494820980264}}.
		
		\bibitem{nish09}
		G.~Chowell, J.~Hyman, L.~Bettencourt, C.~Castillo-Chavez, Mathematical and
		Statistical Estimation Approaches in Epidemiology, 1st Edition, Springer,
		2009.
		
		\bibitem{nish20}
		H.~Nishiura, N.~Linton, A.~Akhmetzhanov, Serial interval of novel coronavirus
		({COVID-19}) infections, International Journal of Infectious Diseases 93~(1)
		(2020) 284--286.
		\newblock \href {http://dx.doi.org/10.1016/j.ijid.2020.02.060}
		{\path{doi:10.1016/j.ijid.2020.02.060}}.
		
	\end{thebibliography}

	\section*{Acknowledgements}
	
	This work was supported by National Research and Innovation Agency (BRIN), project
	number 133/FI/P-KCOVID-19 2B3/IX/2020, 835/IT1 B07/KS.00.00/2020. HS is also supported by Khalifa University through a Faculty Start-Up Grant (No.\ 8474000351-FSU-2021-011). RK gratefully acknowledged financial support from Riset Peningkatan Kapasitas ITB 2020. Results from this research have been used by BRIN to monitor COIVD-19 transmission in Indonesia: \url{https://www.brin.go.id/covid19/covid-meter/}.
	
	\section*{Author contributions statement}
	
	A.H. proposed the method and written the codes, H.S., V.T., R.K., E.P., P.H., and N.N. analysed the data, compared the method, and discussed the results. All authors reviewed the manuscript. 
	
	\section*{Competing interests}
	
	The authors declare no competing interests.
\end{document}